# Spin dynamics triggered by sub-terahertz magnetic field pulses

Zhao Wang, Matthäus Pietz, Jakob Walowski,

IV. Physikalisches Institut, Universität Göttingen, Germany

Arno Förster

Fachhochschule Aachen, Germany

Mihail I. Lepsa

Institute für Bio- und Nanosysteme (IBN-1), Forschungszentrum Jülich, GmbH,

Germany

Markus Münzenberg [a]

IV. Physikalisches Institut, Universität Göttingen, Germany

Current pulses of up to 20 A and as short as 3 ps are generated by a low-temperature-grown GaAs (LT-GaAs) photoconductive switch and guided through a coplanar waveguide, resulting in a 0.6 tesla sub-terahertz (THz) magnetic field pulse. The pulse length is directly calibrated using photocurrent autocorrelation. Magnetic excitations in Fe microstructures are studied by time-resolved Kerr spectroscopy. An ultra-fast response time (within less than 10 ps of the magnetization) to the sub THz electromagnetic field pulse is shown.

PACS: 75.30.Ds, 72.25.Rb, 73.50.Fq

a)  *Corresponding author: Markus Münzenberg; e-mail  mmuenze@gwdg.de*

# I. Introduction

There are different ways to drive magnetization dynamics to the limits: in the time domain, the most prominent are femtosecond all-optical excitations [1-3] and field pulse excitations [4-7]. In the first case the time scales are extremely short (~ ps) [8], but the direction of the excitation can not be controlled. In general, the ultrafast perturbation by the femtosecond laser pulse generates a broad spectrum of excitations from high energy (high k-vector) to low energy modes of the coherent precession (k=0) [9].

Øersted field pulses are generally limited to field strength (~ a few mT) and temporal resolution (>30 ps) since they are restricted to the capabilities of high frequency electronics. The record is held by an alternative cost-intensive approach: the generation of a magnetic field pulse by relativistic electron bunches. At the Stanford Linear Accelerator (SLAC), the magnetic field yields up to more than 5 teslas in amplitude and less than a picosecond in pulse length [10, 11]. From the load of the ultrafast and strong field pulses a fracture of the magnetization is observable. Tudosa et al. therefore postulate a limit for the fastest switching of a recording media determined by the magnetization break-up and driven by the intrinsic non-linearity of the Landau-Lifshitz-Gilbert (LLG) equation. Random thermal fluctuations, always present in the magnetic system, are amplified by the driving field pulse [11]. However, on the other hand, to make an electronic device spin ultra fast, a field of about 10 teslas is needed in order to switch the magnetization within a picosecond.

In the following, we shall present an on-chip geometry approach which uses optical switches as a source of picosecond and high-power current pulses to drive the magnetization dynamics towards a similar value range. The transient magnetic field is generated by a photoconductive (or Auston) switch [12] and the magnetization dynamics are probed with a delay by a probe pulse via the magneto-optic Kerr effect (MOKE) as shown in Fig. 1b. The method, named magneto-optic sampling, has been intensively developed by M. R. Freeman over the last few years [4, 13] and allows the observation of the magnetization transient directly in time. Because of the ultrashort carrier lifetime, low-temperature-grown GaAs (LT-GaAs) is of special interest for applications up to THz bandwidths and is widely used [14]. Here we connect both techniques, magneto-optic sampling and THz pulse generation, to establish a "SLAC" on-chip. The process is as follows: first the THz-current pulse is characterized by a photocurrent autocorrelation technique. Then the magnetic response of an Fe stripe to the sub-THz field pulse, experimentally determined by magneto-optic sampling, is given.

**II. Experiment**

In the following the preparation details and dimensions of the on-chip devices are given. The photoconductive switches are prepared by optical lithography on a 1 µm thick LT-GaAs film grown by molecular beam epitaxy (MBE) on a semi-insulating GaAs wafer at 200 °C and annealed at 600 °C for 10 minutes inside the chamber in As-rich conditions [15]. Characterization of the photo-carrier lifetimes by time-resolved reflectivity measurements reveal two dominating relaxation times of the carriers of 70 fs and 140 fs respectively. In the next step, by using optical lithography, a 22.5 µm wide center conductive strip (5 nm Ti/ 30 nm Al) with a gap of 3 µm is evaporated onto the LT-GaAs substrate. Fig. 1a shows a scanning electron microscope image of the metal-semiconductor-metal (MSM) gap and its dimensions. In addition the experimental geometry is given schematically for the photocurrent autocorrelation experiment. Two pulses delayed by a time $\tau$ illuminate the 3 µm MSM gap and the photocurrent is determined. The electrical pulses are generated by the femtosecond laser illumination of the MSM gap and then transmitted through the coplanar waveguide (Fig 1b). When passing the coplanar waveguide, an ultrafast magnetic field pulse is generated with a dominating in-plane component in the middle of the center conductor.

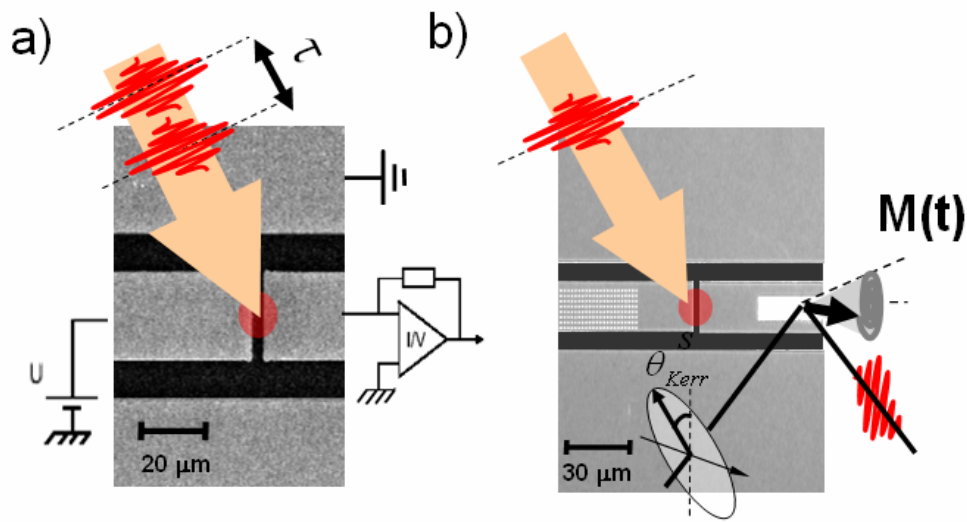

Fig. 1: (color online) a) Scanning Electron microscope image of the optical switch with schematic representation of the photocurrent autocorrelation experiment to determine the current pulse characteristics of the LT-GaAs photo switch. b) Optical microscope image of the optical switch area including the patterned magnetic Fe structures at both sides: an array of two micron-sized structures to the left and an Fe stripe pattern to the right. On top, the schematic representation of the experiment to monitor the magnetization dynamics is given.

To complete the magneto-optic sampling device, the magnetic structures are patterned directly on top of the THz waveguide close to the MSM gap (on-chip geometry) as seen in the optical microscope image shown in Fig. 1b: a 5 nm MgO/ 30 nm Fe film is evaporated on the center conductor and structured using electron-beam lithography by a lift-off process. The Fe structure similar to the one discussed in Section III B can be seen in Fig. 1b, 20 μm × 100 μm in size, close to the MSM gap. On top of the microscopy image, a schematic drawing of the magneto-optic sampling experiment is given. While the first pulse still illuminates the MSM gap, the second pulse is reflected at the magnetic structure. Via the magneto-optic Kerr effect (MOKE), the magnetization dynamics are determined at a delay time $\tau$ using a double modulation technique [16]. Since the skin depth for an Fe film is 3.5 nm at 1 THz frequency only, a considerable current flow through the Fe film itself has to be avoided by the insertion of a thin insulating 5 nm MgO layer. The laser system used for the carrier excitation is a Ti:Sapphire oscillator with a RegA amplifier that generates 60 fs pulses (~ 1 μJ) with a central wavelength of 800 nm and a repetition rate of $f_{rep}$ = 250 kHz.

**III. Experimental results**

**A. Current pulse characteristics**

The advantage of the photocurrent autocorrelation technique presented here is that as opposed to other techniques (e. g. picosecond electro-optic [17] or photoconductive sampling using a dual photoconductor circuit [18]) the same sample geometry as for the magneto-optic sampling can be used to charcterize the electric pulse length. Only a single photoconductor is needed for the photocurrent autocorrelation measurement. A prerequisite is that the photocurrent increases non-linearly with the rise of laser power as seen in Fig. 2a: at a constant voltage, the photocurrent saturates for high fluence. Because of the high defect density of the of LT-GaAs film, the MSM contact has ohmic-like characteristics [15, 19]. It has been shown in [20] that from the photocurrent autocorrelation experiments the time dependent carrier density can be extracted. Therefore the photocurrent autocorrelation curve can be analyzed using an exponential decay function where the time constants are related to carrier relaxation times. In the following we allow two relaxation times ($\tau_{el}$ and $\tau_{geom}$) to describe the experimental data − then the photocurrent as a function of the delay time $\tau$ between the laser pulses is given by:

$$I(\tau) = I_0 - I_{el} e^{-\frac{|\tau|}{\tau_{el}}} - I_{geom} e^{-\frac{|\tau|}{\tau_{geom}}} \qquad (1)$$

where $I_0$ is the maximum photocurrent (Fig. 2b). The parameter set $I_{el}$, $\tau_{el}$ and $I_{geom}$, $\tau_{geom}$ characterize the electrical pulse decay. It is found that the first relaxation time of $\tau_{el}$=1-1.5 ps is related to the carrier recombination time. The ratio of the current amplitudes is about $I_{el} : I_{geom} > 1.5 : 1$. For a finger-switch geometry, where the gap region is curved in order to increase the optically active area, the second, slower decay ($\tau_{geom}$=5-25 ps, dependent on the alignment) can be suppressed. Therefore from the geometry dependence we conclude that antenna effects of the metallization interacting with the fs-light pulse are responsible for the second, slower contribution [21].

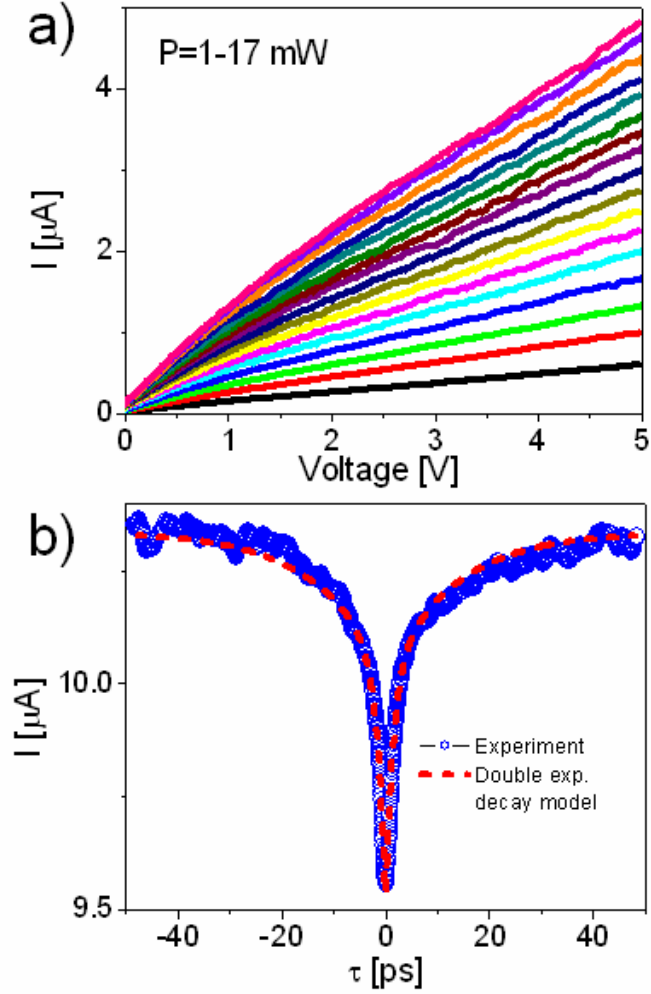

Fig. 2: (color online) a) Current versus voltage characteristics of the photo-switch structure (3 μm gap) under illumination varying the laser power from 1 to 17 mW, showing the non-linearity of the photocurrent with the illumination power. b) Photo-current autocorrelation (3 μm gap structure, 3 V gap voltage). The solid line shows the analysis using a double exponential decay of the photocurrent towards zero delay τ between the two laser pulses illuminating the gap.

The average pulse length for the 3 μm gap switch geometry extracted from the photocurrent autocorrelation experiments is therefore estimated to be $\bar{\tau}$ =3±1 ps from the autocorrelation experiment. This value is in good agreement with the results from photoconductive sampling experiments using a second photoconductive switch as it was determined earlier [18]. An estimate of the maximum current is given by

$$I_{max} = \frac{I_{average} - I_{dark}}{f_{rep}\bar{\tau}}. \qquad (2)$$

It can be easily seen that a high resistance of the non-illuminated switch is needed to suppress the dark current. The average current is up to 16 µA for an 80 V bias voltage and 6 mW average laser power (250 kHz repetition rate) and results in a current amplitude of $I_{max}$=20±8 A. Assuming a homogeneous current distribution throughout the co-planar wave guide (the skin depth of the center conductor materials at 1 THz is about 100 nm), the numerical calculations of the magnetic field distribution above the center conductor result in a homogeneous field component ($B_y$) parallel to the surface of the center conductor of $B_{max}$ = 0.6±2 T. Compared to prior, standard approaches which commonly use Auston switches or electrical pulsers for magneto-optic sampling and synchrotron-based experiments (and which are becoming increasingly important as a novel tool to image magnetization dynamics) this is a significant increase in magnetic field strength. The out-of-plane field component ($B_z$) has a strong contribution at the edges of the conductor, with opposite sign, but it is zero at the center and will be neglected in the following.

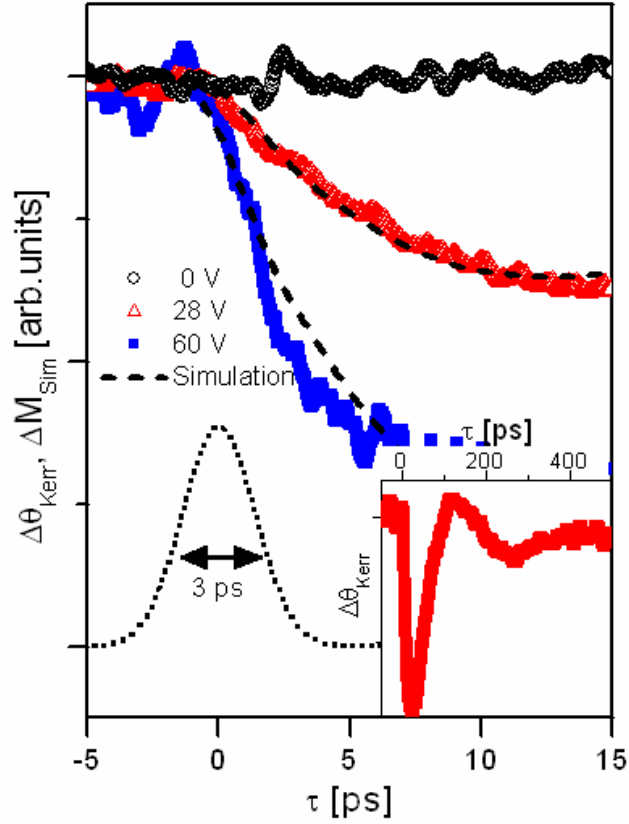

Fig. 3: (color online) Magnetic response of a 30 nm thick Fe stripe pattern on the center conductor for the short time scale for different voltages, 0 V (reference), 28 V and 60 V, applied to the photo-conductive switch. Overlaid on the data, the results of the micro-magnetic simulation are shown (dashed line). In the inset, the signal for 28 V is shown on a larger time scale. As a reference a Gaussian function (3 ps) is plotted to indicate the field pulse (dotted line).

## B. Magnetization dynamics

For the magneto-optic sampling experiments as depicted in Fig. 1b the experimental geometry and a schematic diagram of the sample in the on-chip geometry are shown. An external magnetic static field of 0.03 T is applied along the 20 μm × 100 μm 30 nm Fe film structure to saturate the film along that direction. The Fe film senses the magnetic near field of the THz pulse propagating through the center conductor that is directed perpendicular to the static magnetic field. The evolution of the time resolved Kerr rotation $\Delta\theta_{Kerr}(\tau)$ is shown in Fig. 3. $\Delta\theta_{Kerr}(\tau)$ is probed for 28 V, 60 V and 0 V (reference) voltages applied to the gap. This voltage corresponds to about 0.20±8 T and 0.40±16 T respectively. As a reference a Gaussian function with 3 ps width at half maximum is shown. A steep rising edge of the differential Kerr signal $\Delta\theta_{Kerr}(\tau)$ well below 10 ps is found. This response time does not depend significantly on the field pulse strength. Only the amplitude is about doubled as a result of doubling the field strength. For the reference experiment with zero voltage applied across the photoconductive switch, the observed differential Kerr signal $\Delta\theta_{Kerr}(\tau)$ is zero and thus excludes a direct demagnetization by the laser pump pulse. Micromagnetic simulations using OOMMF [22] represented by the dashed lines are overlaid on the experimental data. For the micromagnetic simulations, an Fe structure of 5 x 15 μm in size and 30 nm in thickness and a cell size of $d_{cell} \leq 50\,nm$ were used. The results were tested for convergence for smaller cell sizes. Surprisingly we find a good agreement with the experimental data without adjusting any parameters (dashed lines in Fig. 3). In the inset of Fig. 3, the Kerr signal shown for the time scale of up to 500 ps (gap voltage 28V) reveals a critically damped oscillation. The high damping found in response to the THz pulse indicates an increase of the apparent damping. This may be interpreted as a signature of the broad spectrum of spin-wave excitations leading to a strong decay of the signal in total. Increasing the field pulse strength can lead to increased damping, as shown in previous cases; the activation of additional damping channels is actually a strongly debated field [23-26] to which we hope to contribute through determining the results of driving the field pulse strength even higher. Details of this increase have yet to be verified in further experiments. The major aim in further experiments will be to realize a full 180-degree switching of the magnetization of the Fe film within one magnetic ultrashort pulse. This will be possible in future photoconductive switch devices approaching 10 T field amplitude.

**IV. Conclusions**

Applying high voltages up to 80 V and an average laser power of 10 mW, the devices are driven to the limit of their stability in the present design. Also low probe beam intensities limit the sensitivity of the Kerr signal detection. However, we have shown that it is possible to generate 0.6±2 T, 3±1 ps long magnetic field pulses and to study the magnetization dynamics excited by a sub-THz electromagnetic field pulse on a chip. The response time of the magnetic signal is found to be within the order of 10 ps, as expected from micro-magnetic calculations. An improved switch design using a finger-switch structure with a larger gap area will stabilize the photoconductive switch and allow pulse strengths of a few teslas (similar to the SLAC experiments, but without using a linear accelerator or synchrotron) in an on-chip experiment with comparatively simple laboratory environment in the future. We expect to study similar effects to these driving the THz radiation emission in ultrafast demagnetization experiments [27] using devices with pulse-rise times below the picosecond range in the future.

**Acknowledgements**

Support by the Deutsche Forschungsgemeinschaft within the priority program SPP 1133 is gratefully acknowledged.